\documentstyle[graphicx,amsmath,amsfonts,mathrsfs,bbm]{article}
\def\e{{\rm e}}
\def\Exp{{\rm Exp}}
\def\exp{{\rm exp}}
\def\d{{\rm d}}

\def\Tr{{\rm Tr}}

\def\Re{{\rm Re}}

\bibliography{plain}
\title{Parametrization of projector-based witnesses for bipartite systems} \vspace{20mm}

\author{
 S. J. Akhtarshenas\thanks{E-mail: akhtarshenas@phys.ui.ac.ir}
 , A. Kheirollahi\thanks{E-mail: a.kheirollahi@sci.ui.ac.ir}
\\
{\small Department of Physics, University of Isfahan, Isfahan, Iran } \\
{\small Quantum Optics Group, University of Isfahan, Isfahan, Iran
}}
\begin{document}
\maketitle

\begin{abstract}
Entanglement witnesses are nonpositive Hermitian operators which
can detect the presence of entanglement. In this paper, we provide
a general parametrization for orthonormal basis of ${\mathbb C}^n$
and use it to construct projector-based witness operators for
entanglement detection in the vicinity of pure bipartite states.
Our method to parameterize entanglement witnesses is operationally
simple and  could be used for doing symbolic and numerical
calculations. As an example
 we use the method for detecting entanglement between an atom and
the single mode of quantized field, described by  the
Jaynes-Cummings model.  We also compare the detection of witnesses
with the negativity of the state, and show that in the vicinity of
pure stats such constructed witnesses able to detect entanglement
of the state.


\end{abstract}


\section{Introduction}
The interest on quantum entanglement has dramatically increased
over the last two decades due to the emerging field of quantum
information theory. It turns out that quantum  entanglement
provides a fundamental potential resource for communication and
information processing \cite{ben1,ben2,ben3}. A pure quantum state
of two or more subsystems is said to be entangled if it is not a
product of states of each components. On the other hand, a
bipartite mixed state $\rho$ is said to be entangled if it can not
be expressed as a convex combination of pure product states
\cite{werner}, otherwise, the state is separable or classically
correlated. It is, therefore, of primary importance testing
whether a given state is separable or entangled. For systems with
dimensions $2\otimes 2$ or $2\otimes 3$, there exists an
operationally simple necessary and sufficient condition for
separability, the so called Peres-Horodecki criterion
\cite{peres,horo0}. It indicates that a state $\rho$ is separable
if and only if the matrix obtained by partially transposing the
density matrix $\rho$ is still positive. However, in higher
dimensional systems this is only a necessary condition; that is,
there exist entangled states whose partial transpose is positive.

Peres-Horodecki criterion for separability leads to a natural
computable measure of entanglement, called negativity
\cite{zycz1,zycz2,vidal}. Negativity is based on the trace norm of
the partial transpose $\rho^{T_1}$ of the bipartite mixed state
$\rho$, and measures the degree to which $\rho^{T_1}$ fails to be
positive, i.e. the absolute value of the sum of negative
eigenvalues of $\rho^{T_1}$
\begin{equation}
{\mathcal N}(\rho)\equiv \frac{\parallel
\rho^{T_1}\parallel_1-1}{2},
\end{equation}
where $\parallel \rho^{T_1}\parallel_1$ denotes the trace norm of
$\rho^{T_1}$. Vidal and Werner \cite{vidal} proved that the
negativity ${\mathcal N}(\rho)$ is an entanglement monotone and
therefore it is a good measure of entanglement.

The most general approach for detecting entanglement is using
entanglement witnesses \cite{horo0,terhal1,terhal2,terhal3,lewen}.
Entanglement witnesses are operators that are designed directly
for distinguishing between separable and entangled states. By
definition, we say that a Hermitian operator $W$ defined on the
product space ${\mathcal H}={\mathcal H}_1\otimes {\mathcal H}_2$
is an entanglement witness if and only if: 1) $\Tr(W \sigma)\ge 0$
for all separable states $\sigma\in {\mathcal S}$, and 2) there
exists at least one entangled state $\rho$ such that $\Tr(W \rho)<
0$. The negative expectation value is hence a signature of
entanglement, and for a state $\rho$ with $\Tr(W \rho)< 0$ we say
that it is detected by $W$. It turns out that a state is entangled
if and only if it is detected by some entanglement witnesses $W$
\cite{horo0}.

An important class of entanglement witnesses is the so called
projector-based witness. Given a pure entangled state
$|\psi\rangle\in{\mathcal H}_1\otimes {\mathcal H}_2$, its
entanglement witness is given by
\begin{equation}\label{W}
W_{\psi}=k(\psi) \mathbbm{1}-|\psi\rangle\langle \psi|,
\end{equation}
where $k(\psi)$ comes from the maximal fidelity between
$|\psi\rangle$ and a product state, i.e.
\begin{equation}\label{k}
k(\psi)=\max_{|e,f\rangle \in \mathcal{S}}|\langle
e,f|\psi\rangle|^2.
\end{equation}

In this paper, we provide an explicit parametrization for the
general orthonormal basis of the Hilbert space ${\mathbb C}^n$.
Naturally, such a parametrization is closely related to the
parametrization of unitary matrices $U(n)$ \cite{Dita2005}, and
therefore such basis requires, in general, $n^2$ real parameters,
i.e. the dimension of unitary group $U(n)$. This parametrization
can be useful in problems arising in quantum information theory.
For instance,  they can be used to construct maximally entangled
states (or generalized Bell states) of a bipartite system, or
Greenberger-Horne-Zeilinger (GHZ) \cite{GHZ} states of a
multiqubit system. We use these maximally entangled states and
construct projector-based witnesses for detecting entanglement.
Our method to construct entanglement witnesses is operationally
simple, in the sense that they can be stored in a computer and
that could be used for doing symbolic and numerical calculations.
As an example we use the method for detecting entanglement between
an atom and the single mode of quantized field, described by  the
Jaynes-Cummings model \cite{JC}.  We also compare the detection of
witnesses with the negativity of the state, and show that in the
vicinity of pure stats such constructed witnesses able to detect
entanglement of the state.

The organization of the paper is as follows. In section 2 we
provide an explicit parametrization for the general orthonormal
basis of the Hilbert space ${\mathbb C}^n$. In section 3 we
construct projector-based entanglement witnesses. In section 4 we
start by reviewing the Jaynes-Cummings model and its solutions and
calculate the negativity of the final state and use the
constructed witnesses to check its separability.  The paper is
concluded in section 5 with a brief conclusion.

\section{Parametrization of the general orthonormal basis}
The aim of this section is to introduce the most general
orthonormal basis ${\mathcal B}$ for ${\mathbb C}^{n}$. We denote
this basis by
\begin{equation}
{\mathcal B}=\{|\Phi^{(0)}\rangle, |\Phi^{(1)}\rangle,
|\Phi^{(2)}\rangle,\cdots,|\Phi^{(n-1)}\rangle\}.
\end{equation}
Let $\{|e_i\rangle\}_{i=0}^{n-1}$ be the  computational
orthonormal basis for ${\mathbb C}^n$. In this basis a general
normalized vector $|\Phi^{(0)}\rangle$  can be expressed by $2n-1$
real parameters as (we do not remove  the total phase)
\begin{eqnarray}\label{Phi0}\nonumber
|\Phi^{(0)}\rangle&=&\cos{\theta_1^{(0)}}\;\e^{i\phi_0^{(0)}}|e_{0}\rangle
\\ \nonumber
&+&\sin{\theta_1^{(0)}}\cos{\theta_2^{(0)}}\;\e^{i\phi_1^{(0)}}|e_{1}\rangle \\
&+&\cdots \\ \nonumber
&+&\sin{\theta_1^{(0)}}\cdots\sin{\theta_{n-2}^{(0)}}\cos{\theta_{n-1}^{(0)}}\;\e^{i\phi_{n-2}^{(0)}}|e_{n-2}\rangle
\\
\nonumber
&+&\sin{\theta_1^{(0)}}\cdots\sin{\theta_{n-2}^{(0)}}\sin{\theta_{n-1}^{(0)}}\;\e^{i\phi_{n-1}^{(0)}}|e_{n-1}\rangle.
\end{eqnarray}
Also we can find  $n-1$  vectors, which are orthonormal to the
above state, as \cite{Dita2005}
\begin{equation}
|\Phi_{k+1}^{(0)}\rangle=\frac{\d}{\d\theta_k^{(0)}}|\Phi^{(0)}\rangle
\mid_{\theta_1^{(0)}=\theta_2^{(0)}=\cdots=\theta_{k-1}^{(0)}=\pi/2},
\qquad k=1,\cdots,n-1,
\end{equation}
where in the above formula one calculates first the derivative and
afterwards the restriction to $\pi/2$. Despite the fact that the
set $\{|\Phi^{(0)}\rangle,|\Phi_{k+1}^{(0)}\rangle\}_{k=1}^{n-1}$
is orthonormal and constitutes a basis for ${\mathbb C}^n$, but
they are not in general form, in the sense that any $U(n-1)$
transformation of the subset
$\{|\Phi_{k+1}^{(0)}\rangle\}_{k=1}^{n-1}$ is also orthonormal to
$|\Phi^{(0)}\rangle$. We therefore define the new vector
$|\Phi^{(1)}\rangle$ as a  linear combination of all vectors of
the subset $\{|\Phi_{k+1}^{(0)}\rangle\}_{k=1}^{n-1}$ as below
\begin{eqnarray}\nonumber
|\Phi^{(1)}\rangle&=&\cos{\theta_1^{(1)}}\;\e^{i\phi_0^{(1)}}|\Phi_2^{(0)}\rangle \\
\nonumber
&+&\sin{\theta_1^{(1)}}\cos{\theta_2^{(1)}}\;\e^{i\phi_1^{(1)}} |\Phi_3^{(0)}\rangle \\
&+& \cdots \\ \nonumber
&+&\sin{\theta_1^{(1)}}\cdots\sin{\theta_{n-3}^{(1)}}\cos{\theta_{n-2}^{(1)}}\;\e^{i\phi_{n-3}^{(1)}}|\Phi_{n-1}^{(0)}\rangle
\\ \nonumber
&+&\sin{\theta_1^{(1)}}\cdots\sin{\theta_{n-2}^{(1)}}\sin{\theta_{n-2}^{(1)}}\;\e^{i\phi_{n-2}^{(1)}}|\Phi_{n}^{(0)}\rangle.
\end{eqnarray}
Obviously, this vector is orthonormal to the $|\Phi^{(0)}\rangle$.
Furthermore, there exist also $n-2$ vectors orthonormal to
$|\Phi^{(1)}\rangle$ as
\begin{equation}
|\Phi^{(1)}_{k+1}\rangle=\frac{\d}{\d\theta_k^{(1)}}|\Phi^{(1)}\rangle
\mid_{\theta_1^{(1)}=\theta_2^{(1)}=\cdots=\theta_{k-1}^{(1)}=\pi/2},
\qquad k=1,\cdots,n-2.
\end{equation}
By construction, these vectors are also orthonormal to the primary
vector $|\Phi^{(0)}\rangle$. Again this new subset
$\{|\Phi^{(1)}_{k+1}\rangle\}_{k=1}^{n-2}$ is not unique and any
$U(n-2)$ transformation of it, has also the same property.
Therefore the third vector of the set ${\mathcal B}$ can be
obtained by making the  linear combination of the above vectors as
\begin{eqnarray}\nonumber
|\Phi^{(2)}\rangle&=&\cos{\theta_{1}^{(2)}}\;\e^{i\phi_{0}^{(2)}}|\Phi^{(1)}_2\rangle \\
\nonumber
&+&\sin{\theta_{1}^{(2)}}\cos{\theta_{2}^{(2)}}\;\e^{i\phi_{1}^{(2)}} |\Phi^{(1)}_3\rangle \\
&+& \cdots \\ \nonumber
&+&\sin{\theta_{1}^{(2)}}\cdots\sin{\theta_{n-4}^{(2)}}\cos{\theta_{n-3}^{(2)}}\;\e^{i\phi_{n-4}^{(2)}}|\Phi^{(1)}_{n-2}\rangle
\\ \nonumber
&+&\sin{\theta_{1}^{(2)}}\cdots\sin{\theta_{n-3}^{(2)}}\sin{\theta_{n-3}^{(2)}}\;\e^{i\phi_{n-3}^{(2)}}|\Phi^{(1)}_{n-1}\rangle.
\end{eqnarray}
Taking the derivatives of $|\Phi^{(2)}\rangle$ with respect to
$\theta_k^{(2)}$ for $k=1,\cdots,n-3$, and making linear
combination of the obtained vectors, we get the vector
$|\Phi^{(3)}\rangle$.  Continuing this procedure, iteratively, we
can find all elements of the orthonormal basis ${\mathcal B}$,
which in summary can be written as ($m=0,1,\cdots,n-1$)
\begin{eqnarray}\nonumber
|\Phi^{(m)}\rangle&=&\cos{\theta_{1}^{(m)}}\;\e^{i\phi_{0}^{(m)}} |\Phi^{(m-1)}_{2}\rangle \\
\nonumber
&+&\sin{\theta_{1}^{(m)}}\cos{\theta_{2}^{(m)}}\;\e^{i\phi_{1}^{(m)}} |\Phi^{(m-1)}_{3}\rangle \\
&+& \cdots \\ \nonumber
&+&\sin{\theta_{1}^{(m)}}\cdots\sin{\theta_{n-m-2}^{(m)}}\cos{\theta_{n-m-1}^{(m)}}\;\e^{i\phi_{n-m-2}^{(m)}}|\Phi^{(m-1)}_{n-m}\rangle
\\ \nonumber
&+&\sin{\theta_{1}^{(m)}}\cdots\sin{\theta_{n-m-1}^{(m)}}\sin{\theta_{n-m-1}^{(m)}}\;\e^{i\phi_{n-m-1}^{(m)}}|\Phi^{(m-1)}_{n-m+1}\rangle,
\end{eqnarray}
where
\begin{equation}
|\Phi^{(m)}_{k+1}\rangle=\frac{\d}{\d\theta_k^{(m)}}|\Phi^{(m)}\rangle
\mid_{\theta_1=\theta_2=\cdots=\theta_{k-1}=\pi/2}, \qquad
k=1,\cdots,n-m-1.
\end{equation}
Here we have defined $|\Phi_{k}^{(-1)}\rangle=|e_{k-2}\rangle$.
From this it is clear that for a given $m$, the number of
parameters required to express $|\Phi^{(m)}\rangle$ as a linear
combination of the vectors of the subset
$\{|\Phi^{(m-1)}_{k+1}\rangle\}_{k=1}^{n-m}$ is equal to
$2(n-m)-1$, and consequently we need, in general,
$\sum_{m=0}^{n-1}(2(n-m)-1)=n^2$ parameters, i.e. $n(n-1)/2$
angles and $n(n+1)/2$ phases. This number is, actually, the
dimension of the group of unitary transformation $U(n)$. Indeed if
we write such constructed basis as
$|\Phi^{(m)}\rangle=\sum_{i=0}^{n-1}U_{mi}|e_i\rangle$,
($m=0,1,\cdots,n-1$), then it is this matrix $U$ which is unitary
with
$\det{U}=\Exp{\left(i\sum_{m=0}^{n-1}\sum_{i=1}^{n-m-1}\phi_{i}^{(m)}\right)}$.
It follows therefore that the $SU(n)$ transformation can be
achieved if we add the requirement
$\sum_{m=0}^{n-1}\sum_{i=1}^{n-m-1}\phi_{i}^{(m)}=0$.

For more illustration of the method we give below two simple
examples. First let us consider $n=2$. In this case  we have
\begin{eqnarray}\label{Psin2}\nonumber
|\Phi^{(0)}\rangle&=&\cos{\theta_1}\;\e^{i\phi_0}|e_1\rangle
+\sin{\theta_1}\;\e^{i\phi_1}|e_2\rangle, \\
|\Phi^{(1)}\rangle
&=&-\sin{\theta_1}\;\e^{(i\phi_0+\xi_0)}|e_1\rangle+\cos{\theta_1}\;\e^{i(\phi_1+\xi_0)}|e_2\rangle.
\end{eqnarray}
Also for $n=3$ we have
\begin{eqnarray}\label{Psin3}\nonumber
|\Phi^{(0)}\rangle&=&\cos{\theta_1}\;\e^{i\phi_0}|e_1\rangle
+\sin{\theta_1}\cos{\theta_2}\;\e^{i\phi_1}|e_2\rangle
+\sin{\theta_1}\sin{\theta_{2}}\;\e^{i\phi_{2}}|e_3\rangle,
\\ \nonumber
|\Phi^{(1)}\rangle&=&-\cos{\eta_1}\sin{\theta_1}\e^{i(\phi_0+\xi_0)}|e_1\rangle \\
\nonumber
&+&\left(\cos{\eta_1}\cos{\theta_1}\cos{\theta_2}\;\e^{i(\xi_0+\phi_1)}-\sin{\eta_1}\sin{\theta_2}\;\e^{i(\xi_1+\phi_1)}\right)|e_2\rangle
\\ \nonumber
&+&\left(\cos{\eta_1}\cos{\theta_1}\sin{\theta_{2}}\;\e^{i(\xi_0+\phi_{2})}+\sin{\eta_1}\cos{\theta_{2}}\;\e^{i(\xi_1+\phi_{2})}\right)|e_3\rangle,
\\ \nonumber
|\Phi^{(2)}\rangle&=&\sin{\eta_1}\sin{\theta_1}\e^{i(\zeta_0+\phi_0+\xi_0)}|e_1\rangle \\
\nonumber
&-&\left(\sin{\eta_1}\cos{\theta_1}\cos{\theta_2}\;\e^{i(\zeta_0+\xi_0+\phi_1)}
+\cos{\eta_1}\sin{\theta_2}\;\e^{i(\zeta_0+\xi_1+\phi_1)}\right)|e_2\rangle
\\
&+&\left(-\sin{\eta_1}\cos{\theta_1}\sin{\theta_{2}}\;\e^{i(\zeta_0+\xi_0+\phi_{2})}
+\cos{\eta_1}\cos{\theta_{2}}\;\e^{i(\zeta_0+\xi_1+\phi_{2})}\right)|e_3\rangle,
\end{eqnarray}
where for the sake of simplicity we have used
$\theta_i^{(0)}=\theta_i,\;\phi_i^{(0)}=\phi_i,\;
\theta_i^{(1)}=\eta_i, \; \phi_i^{(1)}=\xi_i, \;
\phi_0^{(2)}=\zeta_0$.

\section{Entanglement witnesses}
Now in this section we attempt to use the above parametrization to
construct entanglement witnesses.   Given a pure entangled state
$|\psi\rangle$ belongs to ${\mathbb C}^{n_1}\otimes {\mathbb
C}^{n_2}$, its entanglement witness is given by
\begin{equation}\label{W}
W_{\psi}=k(\psi) \mathbbm{1}-|\psi\rangle\langle \psi|,
\end{equation}
where $k(\psi)$ comes from the maximal fidelity between
$|\psi\rangle$ and a product state, i.e.
\begin{equation}\label{k}
k(\psi)=\max_{|e,f\rangle \in {\mathcal S}}|\langle
e,f|\psi\rangle|^2,
\end{equation}
where ${\mathcal S}$ denotes the set of all separable states. This
entanglement witness detects entanglement around the entangled
state $|\psi\rangle$. In general, it is not easy to calculate the
constant $k(\psi)$, except for a two-qubit system  that there
exists a simple relation as
\begin{equation}\label{kpsi}
k(\psi)=\frac{1}{2}\left(1+\sqrt{1-(C(\psi))^{2}}\right),
\end{equation}
where $C(\psi)$ is the so called Wootters concurrence
\cite{wootters}, and for a pure state
$|\psi\rangle=a_{00}|00\rangle+a_{01}|01\rangle+a_{10}|10\rangle+a_{11}|11\rangle$,
has the  form $C(\psi)=2|a_{00}a_{11}-a_{01}a_{10}|$. It is clear
from equation (\ref{kpsi}) that $k(\psi)$ ranges from 1 to
$\frac{1}{2}$  as $C(\psi)$ goes from 0 to 1, so that the minimum
value for this constant happens whenever $|\psi\rangle$ is a
maximally entangled state, i.e. a Bell state. On the other hand
for a general bipartite system it is shown that the constant
$k(\psi)$ equals  to the square of the  maximal Schmidt number of
the state $|\psi\rangle$ \cite{Mohamed2004}. According to the
Schmidt theorem, any bipartite pure state $|\psi\rangle
\in{\mathbb C}^{n_1}\otimes {\mathbb C}^{n_2}$ can be written in
the following form \cite{Schmidt}
\begin{equation}\label{SchmidtD}
|\psi\rangle=\sum_{i=1}^{n}\sqrt{\lambda_i}|u_i\rangle|v_i\rangle,
\end{equation}
with $1\le n\le \min{(n_1,n_2)}$ and $\lambda_i\ge 0$ with
$\sum_{i=1}^{n}\lambda_i=1$, and where $\{|u_i\rangle\}_{i=1}^{n}$
and $\{|v_i\rangle\}_{i=1}^{n}$ are the orthonormal eigenvectors
of the reduced density operators
$\rho_{1}=\rm{Tr}_{2}\left(|\psi\rangle\langle\psi|\right)$ and
$\rho_{2}=\rm{Tr}_{1}\left(|\psi\rangle\langle\psi|\right)$,
respectively.  The number $n$ is called the Schmidt rank of
$|\psi\rangle$.  As a result of the Schmidt rank, we can say that
if a system has dimension $m$, then it can not be entangled with
more than $m$ orthogonal states of the another system.

Motivated by this, we now use the orthonormal basis  ${\mathcal
B}$ and ${\mathcal B}^\prime$ for the Hilbert spaces ${\mathbb
C}^{n_1}$ and ${\mathbb C}^{n_2}$ respectively, and define  a
bipartite pure state $|\Psi^{[n]}\rangle$ in Schmidt form as
\begin{eqnarray}\label{Psi}\nonumber
|\Psi^{[n]}\rangle&=&\cos{\alpha_1}|\Phi^{(0)}\rangle|{\Phi^\prime}^{(0)}\rangle
\\ \nonumber
&+&\sin{\alpha_1}\cos{\alpha_2}|\Phi^{(1)}\rangle|{\Phi^\prime}^{(1)}\rangle \\
&+&\cdots \\ \nonumber
&+&\sin{\alpha_1}\cdots\sin{\alpha_{n-2}}\cos{\alpha_{n-1}}|\Phi^{(n-2)}\rangle|{\Phi^\prime}^{(n-2)}\rangle
\\
\nonumber
&+&\sin{\alpha_1}\cdots\sin{\alpha_{n-2}}\sin{\alpha_{n-1}}|\Phi^{(n-1)}\rangle|{\Phi^\prime}^{(n-1)}\rangle.
\end{eqnarray}
Obviously, the generalized Bell states obtained whenever all
Schmidt numbers become $\frac{1}{\sqrt{n}}$, which occurs when
$\alpha_i=\cos^{-1}{\frac{1}{\sqrt{n-i+1}}}$ for $i=1,\cdots,n-1$.
We can, therefore, define the witness operator based on the pure
state $|\Psi^{[n]}\rangle$  as
\begin{equation}\label{WPsi}
W_{\Psi^{[n]}}=k(\Psi^{[n]}){\mathbbm 1}-|\Psi^{[n]}\rangle\langle
\Psi^{[n]}|,
\end{equation}
where $k(\Psi^{[n]})$ is equal to the square of the maximal
Schmidt number of the ket $|\Psi^{[n]}\rangle$ \cite{Mohamed2004}.
Equation (\ref{k}) guarantees that $\Tr\left(W_{\Psi^{[n]}}
\sigma\right)=k(\Psi^{[n]})-\langle \Psi^{[n]}|\sigma
|\Psi^{[n]}\rangle\ge 0$ for all separable states $\sigma\in
{\mathcal S}$. On the other hand since the Schmidt rank of every
bipartite entangled pure state is greater than one, so all Schmidt
numbers of the entangled bipartite pure states are less than 1 and
therefore $\Tr\left(W_{\Psi^{[n]}} |\Psi^{[n]}\rangle\langle
\Psi^{[n]}|\right)=k(\Psi^{[n]})-1< 0$ for every entangled state
$|\Psi^{[n]}\rangle$, i.e. $W_{\Psi^{[n]}}$ detects entanglement
of the pure state $|\Psi^{[n]}\rangle$.

Now let us consider the state $\rho$ acting on the Hilbert space
${\mathbb C}^{n_1}\otimes {\mathbb C}^{n_2}$. Equation
(\ref{WPsi}) guarantees that if $\langle
W_{\Psi^{[n]}}\rangle_{\rho}=\Tr(W_{\Psi^{[n]}}\rho)=k(\Psi^{[n]})-F_{\Psi^{[n]}}(\rho)<0$,
where $F_{\Psi^{[n]}}(\rho)=\langle
\Psi^{[n]}|\rho|\Psi^{[n]}\rangle$ is the fidelity between two
states  $|\Psi^{[n]}\rangle$ and $\rho$, then $\rho$ is not
separable and has some entanglement. Therefore for a given state
$\rho$ acting on the Hilbert space ${\mathbb C}^{n_1}\otimes
{\mathbb C}^{n_2}$, we say that $W_{\Psi^{[n]}}$ detects
entanglement of $\rho$ if and only if the fidelity between
$|\Psi^{[n]}\rangle$ and $\rho$ is greater than $k(\Psi^{[n]})$,
otherwise $\rho$ is unentangled or its entanglement can not be
detected by $W_{\Psi^{[n]}}$.

\section{ Witnessing entanglement of the Jaynes-Cummings model}
In this section we use such constructed witnesses to detect
entanglement of the Jaynes-Cummings model.  The Jaynes-Cummings
Hamiltonian between a two-level atom $A$ and a single-mode
quantized radiation field $F$ is described by
\begin{equation}\label{JCH}
H=\frac{1}{2}\hbar \omega _A \sigma _z +\hbar \omega _F a^\dag
a+\hbar g(\sigma_{+} \otimes a   + \sigma_{-}   \otimes a^\dag ).
\end{equation}
This Hamiltonian  acts on the product Hilbert space ${\mathcal
H}^A\otimes{\mathcal H}^F$.  Here $g$ is the atom-field coupling
constant, $\omega_A=(\epsilon_e-\epsilon_g)/\hbar$ is the atomic
transition frequency, and $\omega_F$ denotes the field frequency.
The atomic ``spin-flip'' operators $\sigma_{+}=|e\rangle\langle
g|$, $\sigma_{-}=|g\rangle\langle e|$, and the atomic inversion
operator $\sigma_z=|e\rangle\langle e|-|g\rangle\langle g|$ act on
the atom Hilbert space ${\mathcal H}^A={\mathbb C}^2$ spanned by
the excited state $|e\rangle\rightarrow(1,0)^T$ and the ground
state $|g\rangle\rightarrow(0,1)^T$. The field annihilation and
creation operators $a$ and $a^\dag$ satisfy the commutation
relation $[a,a^{\dag}]=1$ and act on the field Hilbert space
${\mathcal H}^F$ spanned by the photon-number states
$\{|n\rangle=\frac{(a^\dag)^n}{\sqrt{n!}}|0\rangle\}_{n=0}^{\infty}$.
In the rest of this section, we consider the solutions of this
Hamiltonian in two cases.

\subsection{Case 1}
We first assume that the atom is initially prepared in the excited
state $\rho^A(0)=|e\rangle\langle e|$, and the field is initially
in the number state $\rho^F(0)=|n\rangle\langle n|$. We also
consider the effect of pure phase decoherence on the
Jaynes-Cummings model. In this situation, the master equation
governing the time evolution for the system under the Markovian
approximation is given by \cite{Gardiner1991}
\begin{equation}\label{1}
\frac{\d\rho}{\d t}=-i[H,\rho]-\frac{\gamma}{2}[H,[H,\rho]],
\end{equation}
where $\gamma$ is the phase decoherence coefficient. The formal
solution of this equation can be expressed as
\begin{equation}\label{2}
\rho(t)=\sum\limits_{k=0}^{\infty}\frac{(\gamma
t)^{k}}{k!}M^{k}(t)\rho(0)M^{\dag k}(t),
\end{equation}
where $\rho(0)=\rho^A(0)\otimes \rho^F(0)$ is the initial state of
the system and $M^k(t)$ is defined by
\begin{equation}\label{3}
M^{k}(t)=H^{k}\;\exp{(-iHt)}\;\exp{(\frac{-\gamma t}{2}H^{2})}.
\end{equation}
 Then the time evolution of the system reads
\cite{Li2003}
\begin{eqnarray}\label{4}
\nonumber
 \rho_{n}(t)&=&E_{n}|e\rangle\langle
e|\otimes|n\rangle\langle n|
+G_{n}|e\rangle\langle g|\otimes|n\rangle\langle n+1| \\
&+& G^\ast_{n}|g\rangle\langle e|\otimes|n+1\rangle\langle
n|+F_{n}|g\rangle\langle g|\otimes|n+1\rangle\langle n+1|,
\end{eqnarray}
where we have defined
\begin{eqnarray}\nonumber
E_{n}&=&\frac{1}{4}\left(2+\frac{\Delta^{2}}{2\Omega_{n}^{2}}+(2-\frac{\Delta^{2}}{2\Omega_{n}^{2}})\cos{2\Omega_{n}t}\;\exp{(-2\gamma
t \Omega_{n}^{2})}\right),\\\nonumber
F_{n}&=&\frac{1}{4}\frac{g^{2}(n+1)}{\Omega_{n}^{2}}\left(2-2\cos{2\Omega_{n}t}\;\exp{(-2\gamma
t \Omega_{n}^{2})}\right),\\\nonumber
G_{n}&=&\frac{g\sqrt{(n+1)}}{4\Omega_{n}}\left(\frac{\Delta}{\Omega_{n}}\left(1-\cos{2\Omega_{n}t}\;\exp{(-2\gamma
t \Omega_{n}^{2})}\right)+2i\sin{2\Omega_{n}t}\;\exp{(-2\gamma t
\Omega_{n}^{2})}\right),
\end{eqnarray}
where $\Delta=\omega_{A}-\omega_{F}$ ,
$\Omega_{n}=\sqrt{\frac{\Delta^{2}}{4}+g^{2}(n+1)}$. The
negativity of this state is easily obtained as
\begin{equation}\label{9}
{\mathcal
N}(\rho_{n})=\frac{g\sqrt{n+1}}{2\Omega_{n}}\sqrt{\left(\frac{\Delta^{2}}{\Omega_{n}^{2}}\left(1-\exp{(-2\gamma
t\Omega_{n}^{2} )}\cos{2\Omega_{n}t}\right)^{2}+4\;\exp{(-4\gamma
t\Omega_{n}^{2} )}\sin^{2}{2\Omega_{n}t}\right)}.
\end{equation}
Now we attempt to characterize the separability of the above state
with witnesses defined in equation (\ref{WPsi}). To this aim  we
first consider a Bell state as
\begin{equation}\label{Psi2}
|\Psi^{[2]}\rangle=\frac{1}{\sqrt{2}}\left(|\Phi^{(0)}\rangle|{\Phi^\prime}^{(0)}\rangle
+|\Phi^{(1)}\rangle|{\Phi^\prime}^{(1)}\rangle\right),
\end{equation}
where $\{|\Phi^{(0)}\rangle,\; |\Phi^{(1)}\rangle\}$ are two
orthonormal  vectors in the field space spanned by
$\{|n\rangle,\;|n+1\rangle\}$, and
$\{|{\Phi^\prime}^{(0)}\rangle,\; |{\Phi^\prime}^{(1)}\rangle\}$
are two orthonormal vectors in the atomic  space spanned by
$\{|e\rangle,\;|g\rangle\}$. The parametrization of both sets of
vectors are given by equation (\ref{Psin2}).  We find for the
fidelity between $\rho_{n}(t)$ and $|\Psi^{[2]}\rangle$
\begin{eqnarray}
F_{\Psi^{[2]}}(\rho_n(t))=\frac{1}{2}\left(|Y_1|^2E_n+|Y_2|^2
F_n\right)+\Re\left(Y_1^\ast Y_2
\;\e^{i(\phi_1-\phi_0+\phi_1^{\prime}-\phi_0^{\prime})}G_n\right),
\end{eqnarray}
where $Y_1$ and $Y_2$ are defined by
\begin{eqnarray} \nonumber
Y_1&=&\cos{\theta_1}\cos{\theta_1^{\prime}}+
\sin{\theta_1}\sin{\theta_1^{\prime}}\;\e^{i(\xi_0+\xi_0^\prime)},
\\ \nonumber
Y_2&=&\sin{\theta_1}\sin{\theta_1^{\prime}}+
\cos{\theta_1}\cos{\theta_1^{\prime}}\;\e^{i(\xi_0+\xi_0^\prime)}.
\end{eqnarray}
Now in order to get maximal fidelity,  we can use the software
MATHEMATICA, and maximize $F_{\Psi^{[2]}}(\rho_n(t))$ with respect
to all parameters of the state $|\Psi^{[2]}\rangle$. In figure (1)
we have plotted negativity ${\mathcal N}(\rho_1(t))$ (upper panel)
and the maximum fidelity $F_{\Psi^{[2]}}(\rho_1(t))$ (lower panel)
as a function of $t$ with  $g=1$, $\gamma=0.3$, $\Delta=1$.  It is
clear from the figure that whenever negativity is nonzero, the
fidelity $F_{\Psi^{[2]}}(\rho_1(t))$ is also greater than $1/2$,
which indicates the detection of entanglement by $W_{\Psi^{[2]}}$.
\begin{figure}[t]
\centerline{\includegraphics[height=8cm]{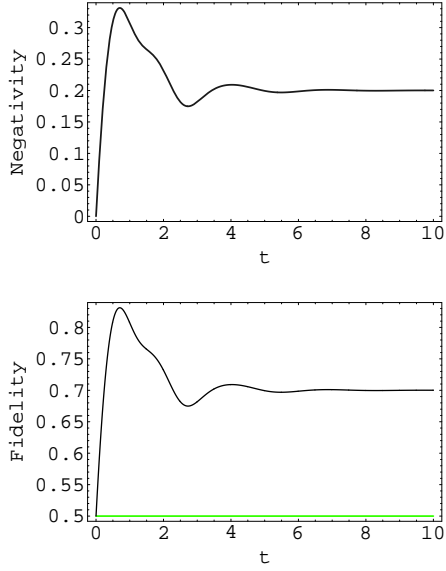}}
\caption{Negativity ${\mathcal N}(\rho_1(t))$ (upper panel) and the
fidelity $F_{\Psi^{[2]}}(\rho_1(t))$ (lower panel) are plotted as a
function of $t$ with  $g=1$, $\gamma=0.3$, $\Delta=1$. The
horizontal line in the lower figure shows the minimum fidelity that
above it the state is detected to be entangled.}
\end{figure}

\subsection{Case 2}
Since, in practice, it is difficult to realize an atom in a pure
state, therefore we now suppose that the atom is prepared,
initially, in a general mixed state  with the diagonal
representation
\begin{equation}
\rho^A(0)= \lambda|g\rangle\langle g | + (1-\lambda)| e \rangle
\langle e |, \qquad \lambda\in[0,1],
\end{equation}
but, however, the field is in a pure number state
\begin{equation}
\rho^F(0)=|n\rangle\langle n|.
\end{equation}
However, we ignore the effect of decoherence and assume that the
evolution of the system is unitary. Accordingly, the final state
of the system can be obtained as \cite{Akhtarshenas,furuichi3}
\begin{eqnarray}\nonumber
\rho_n(t)&=&\lambda \left(|A_n|^2|e\rangle\langle
e|\otimes|n-1\rangle\langle n-1|+A_n B^\ast_n|e\rangle\langle
g|\otimes|n-1\rangle\langle n|\right.
\\ \nonumber&+&\left.A^\ast_nB_n|g\rangle\langle
e|\otimes|n\rangle\langle n-1| +|B_n|^2|g\rangle\langle
g|\otimes|n\rangle\langle n|\right)
\\ \nonumber
&+&(1-\lambda)\left(|C_n|^2|e\rangle\langle
e|\otimes|n\rangle\langle n| +C_n D^\ast_n|e\rangle\langle
g|\otimes|n\rangle\langle
n+1|\right. \\
&+&\left.C^\ast_n D_n|g\rangle\langle e|\otimes|n+1\rangle\langle
n| +|D_n|^2|g\rangle\langle g|\otimes|n+1\rangle\langle
n+1|\right),
\end{eqnarray}
where we have defined
\begin{eqnarray}\nonumber
A_n&=&-i{\rm
e}^{-i\omega_F(n-1/2)t}\frac{g\sqrt{n}}{\Omega_{n-1}}\sin{\Omega_{n-1}t},
\\ \nonumber
B_n&=&{\rm e}^{-i\omega_F(n-1/2)t} \left(\cos{\Omega_{n-1}
t}+i\frac{\Delta}{2\Omega_{n-1}}\sin{\Omega_{n-1} t}\right),
\\ \nonumber
C_n &=&{\rm e}^{-i\omega_F(n+1/2)t}\left(\cos{\Omega_n t
}-i\frac{\Delta}{2\Omega_{n}}\sin{\Omega_{n}t}\right),
\\
 D_n&
=&-i{\rm e}^{-i
\omega_F(n+1/2)t}\frac{g\sqrt{n+1}}{\Omega_{n}}\sin{\Omega_n t},
\end{eqnarray}
where the Rabi frequency $\Omega_n$, and the detuning parameter
$\Delta$ are the same as before. For the above state the
negativity can be expressed as
\begin{eqnarray} \nonumber
{\mathcal
N}(\rho_n(t))&=&\frac{1}{2}\left(\sqrt{\lambda^2|B_n|^4+4(1-\lambda)^2|C_n|^2|D_n|^2}-\lambda|B_n|^2\right)
\\
&+&\frac{1}{2}\left(\sqrt{(1-\lambda)^2|C_n|^4+4\lambda^2|A_n|^2|B_n|^2}-(1-\lambda)|C_n|^2\right).
\end{eqnarray}
\begin{figure}[t]
\centerline{\includegraphics[height=6cm]{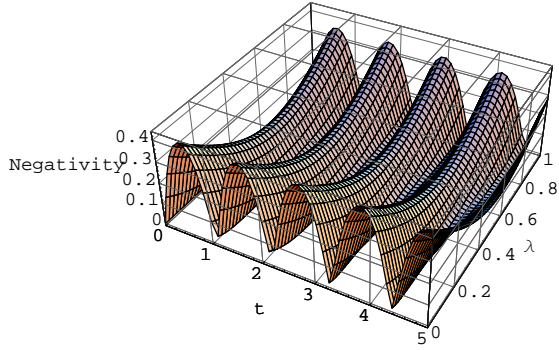}}
\caption{Negativity ${\mathcal N}(\rho_1(t))$ is plotted as a
function of $t$ with  $g=1$,  $\Delta=1$. .}
\end{figure}
In order to show the effect of mixing parameter $\lambda$ on the
negativity of the system, we plot the negativity ${\mathcal
N}(\rho_n(t))$ as a function of $t$ and $\lambda$ with $g=1$,
$\Delta=5$  in figure (2). It is clear from this figure that when
purity of this state is decreased, the negativity is also decreased.

\begin{figure}[t]
\centerline{\includegraphics[height=8cm]{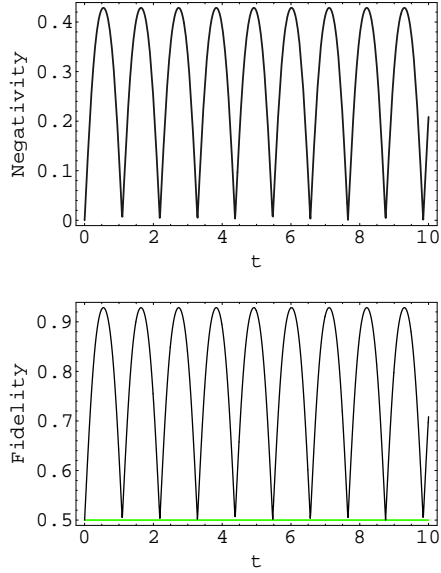}}
\caption{Negativity ${\mathcal N}(\rho_1(t))$ (upper panel) and the
fidelity $F_{\Psi^{[2]}}(\rho_1(t))$ (lower panel) are plotted as a
function of $t$ with  $g=1$, $\Delta=5$ and $\lambda=0$. }
\end{figure}

\begin{figure}[t]
\centerline{\includegraphics[height=8cm]{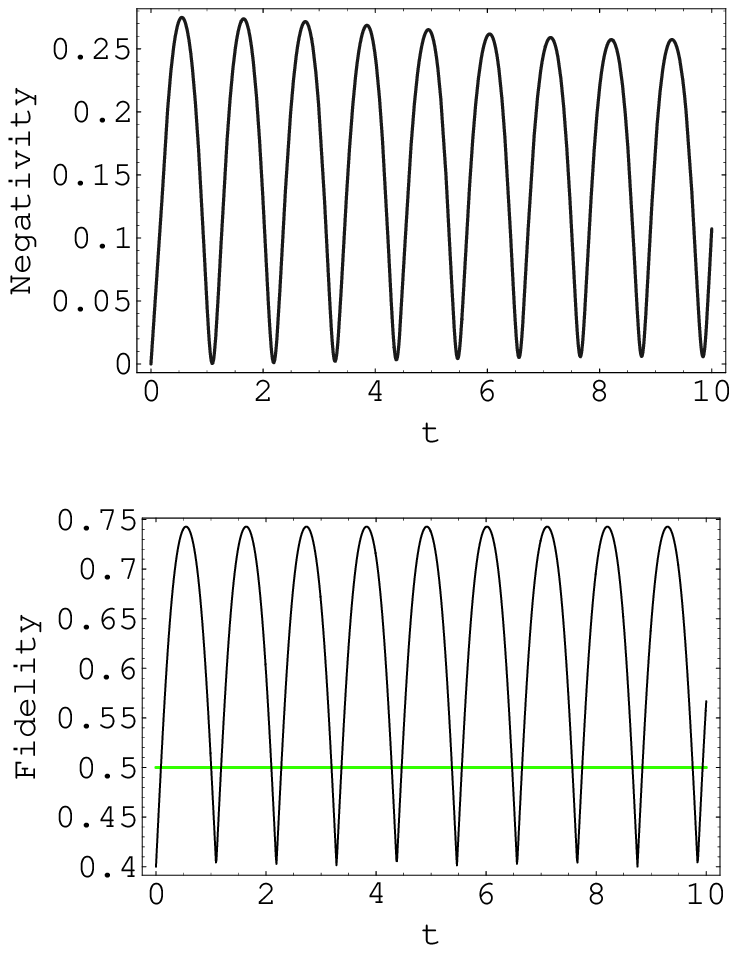}}
\caption{Negativity ${\mathcal N}(\rho_1(t))$ (upper panel) and the
fidelity $F_{\Psi^{[2]}}(\rho_1(t))$ (lower panel) are plotted as a
function of $t$ with  $g=1$, $\Delta=5$ and $\lambda=0.2$. }
\end{figure}
Now in order to characterize entanglement of this state by using
the notion of entanglement witness, we define $|\Psi^{[2]}\rangle$
as equation (\ref{Psi2}), but here  $\{|\Phi^{(0)}\rangle,\;
|\Phi^{(1)}\rangle\}$ are two orthonormal  vectors in the field
space spanned by $\{|n-1\rangle,\; |n\rangle,\;|n+1\rangle\}$ and
$\{|{\Phi^\prime}^{(0)}\rangle,\; |{\Phi^\prime}^{(1)}\rangle\}$
are two orthonormal vectors in the atomic  space spanned by
$\{|e\rangle,\;|g\rangle\}$.  The parametrization of these two
sets of vectors are given by equations (\ref{Psin3}) and
(\ref{Psin2}), respectively. We obtain for the fidelity the
following relation
\begin{eqnarray}
F_{\Psi^{[2]}}(\rho_n(t))&=&\lambda\left(\frac{1}{2}\left(|X_1|^2|A_n|^2+|X_2|^2|B_n|^2\right)+\Re{\left(X_1^\ast
X_2\e^{i(\phi_1-\phi_0+\phi_1^{\prime}-\phi_0^{\prime})}A_n B^\ast_n\right)}\right) \\
\nonumber &+&
(1-\lambda)\left(\frac{1}{2}\left(|X_3|^2|C_n|^2+|X_4|^2|D_n|^2\right)+\Re{\left(X_3^\ast
X_4\e^{i(\phi_2-\phi_1+\phi_1^{\prime}-\phi_0^{\prime})}C_n
D^\ast_n\right)}\right),
\end{eqnarray}
where
\begin{eqnarray} \nonumber
X_1&=&\cos{\theta_1}\cos{\theta_1^{\prime}}+
\cos{\eta}\sin{\theta_1}\sin{\theta_1^{\prime}}\e^{i(\xi_0+\xi_0^\prime)},
\\ \nonumber
X_2&=&\sin{\theta_1}\cos{\theta_2}\sin{\theta_1^{\prime}}+
\left(\cos{\eta}\cos{\theta_1}\cos{\theta_2}\e^{i\xi_0}-\sin{\eta}\sin{\theta_2}\e^{i\xi_1}\right)\cos{\theta_1^{\prime}}\e^{i\xi_0^{\prime}},
 \\ \nonumber
X_3&=&\sin{\theta_1}\cos{\theta_2}\cos{\theta_1^{\prime}}-
\left(\cos{\eta}\cos{\theta_1}\cos{\theta_2}\e^{i\xi_0}-\sin{\eta}\sin{\theta_2}\e^{i\xi_1}\right)\sin{\theta_1^{\prime}}\e^{i\xi_0^{\prime}},
\\
X_4&=&\sin{\theta_1}\sin{\theta_2}\sin{\theta_1^{\prime}}+
\left(\cos{\eta}\cos{\theta_1}\sin{\theta_2}\e^{i\xi_0}+\sin{\eta}\cos{\theta_2}\e^{i\xi_1}\right)\cos{\theta_1^{\prime}}\e^{i\xi_0^{\prime}}.
\end{eqnarray}
Again, by using the MATHEMATICA, we can maximize
$F_{\Psi^{[2]}}(\rho_n(t))$ with respect to all parameters of the
state $|\Psi^{[2]}\rangle$. In figure (3) we have plotted the
negativity ${\mathcal N}(\rho_1(t))$ (upper panel) and the
fidelity $F_{\Psi^{[2]}}(\rho_1(t))$ (lower panel) as a function
of $t$ with  $g=1$, $\Delta=5$ and $\lambda=0$. In this case the
state of the system is pure and it is clear from the figure that
$W_{\Psi^{[2]}}$ enable to detect the entanglement of the system.
On the other hand by increasing $\lambda$ from $0$ to $1/2$, the
purity of the state is decreased, and the ability of
$W_{\Psi^{[2]}}$ to detect entanglement is decreased. For
instance, we plot in figure (4) the negativity and the fidelity as
a function of $t$ for parameters the same as figure (3) but
$\lambda=0.2$. We see that in this case there exist some entangled
states that can not be detected by $W_{\Psi^{[2]}}$.

\section{Conclusion}
We have presented a general parametrization for orthonormal basis
of ${\mathbb C}^n$.  This parametrization can be  used to
construct projector-based witness operators for entanglement
detection in the vicinity of pure multipartite states.  As an
example
 we have used the method for detecting entanglement between an atom and
the single mode of quantized field, described by  the
Jaynes-Cummings model.  We have also compared the detection of
witnesses with the negativity of the state, and have shown that in
the vicinity of pure stats such constructed witnesses able to
detect entanglement of the state.

\newpage

\end{document}